# *RecipeMind*: Guiding Ingredient Choices from Food Pairing to Recipe Completion using Cascaded Set Transformer

Mogan Gim *
Korea University
Seoul, Korea
akim@korea.ac.kr

Donghee Choi *
Korea University
Seoul, Korea
choidonghee@korea.ac.kr

Kana Maruyama
Sony AI
Tokyo, Japan
Kana.Maruyama@sony.com

Jihun Choi
Sony AI
Tokyo, Japan
Jihun.A.Choi@sony.com

Hajung Kim
Korea University
Seoul, Korea
hajungk@korea.ac.kr

Donghyeon Park †
Sejong University
Seoul, Korea
parkdh@sejong.ac.kr

Jaewoo Kang †
Korea University
Seoul, Korea
kangj@korea.ac.kr

## ABSTRACT

We propose a computational approach for recipe ideation, a downstream task that helps users select and gather ingredients for creating dishes. To perform this task, we developed RecipeMind, a food affinity score prediction model that quantifies the suitability of adding an ingredient to set of other ingredients. We constructed a large-scale dataset containing ingredient co-occurrence based scores to train and evaluate RecipeMind on food affinity score prediction. Deployed in recipe ideation, RecipeMind helps the user expand an initial set of ingredients by suggesting additional ingredients. Experiments and qualitative analysis show RecipeMind's potential in fulfilling its assistive role in cuisine domain.

---

* Equal Contributors.
† Corresponding authors.

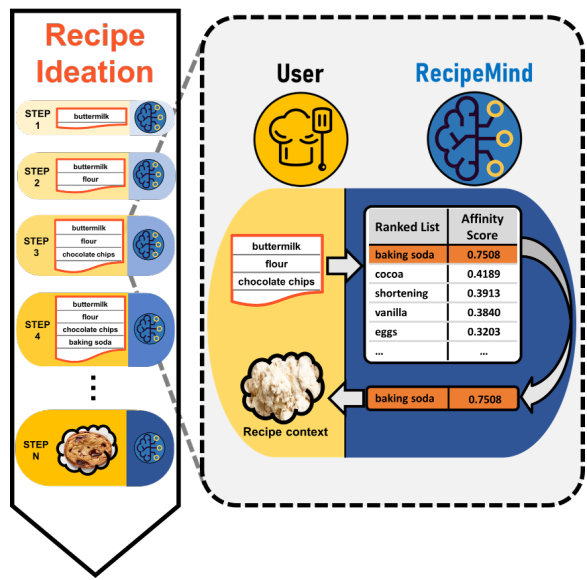

**Figure 1: Overview of Recipe Ideation. RecipeMind guides the user's choices from *food pairing* (Step 1) shown on left side in Recipe Ideation) to *recipe completion* (Step N). In the middle of ideation (Step 3), RecipeMind recommends adding baking soda to current set based on its score predictions.**

## 1 INTRODUCTION

Professional chefs and home cooks have pursued to create new dishes and formulate novel recipe ideas which are important tasks in culinary domain. Some recipe ideas derive from existing dishes while others are created from novel ingredient combinations. As a recipe comprises a set of ingredients and sequence of cooking instructions, one may want to brainstorm a recipe starting with a set of few ingredients and consecutively expanding it with additional ones. As illustrated in Figure 1 starting with **buttermilk**, we define these consecutive steps of selecting ingredients (**flour**, **chocolate chips**, **baking soda**) as *recipe ideation*, gradually leading to recipe completion (*Chocolate Chip Cookies*).

Recipe ideation is challenging due to vast space of cooking possibilities and complexity for flavor yet important for creative cooking in culinary domain [44]. As multiple ingredients used for cooking a dish form recipe context [38], choosing the right ingredient requires comprehensive understanding in culinary aspects such as aroma or flavor [13]. A systematic approach towards recipe ideation would involve initiating with the most basic recipe idea consisting few ingredients and iteratively updating it with new additional ingredients that goes well with its overall recipe context.

Computational approaches for assisting recipe ideation process have recently become necessary to solve these issues in culinary domain [21]. For instance, Kitchenette predicts food pairing scores [40] and RecipeBowl retrieves the best ingredient given a nearly completed recipe [24] which are deemed as earliest and latest stage of recipe ideation respectively. A more versatile computational approach deployable in *any* stages of recipe ideation may be desirable.

In this work, RecipeMind performs the ideation task by quantifying the suitability of adding an ingredient to set of other ingredients. Henceforth, we make the following definitions and formulate the objective of recipe ideation task prior to introducing our proposed model RecipeMind.

- *Ingredient Space* $\mathbb{U}$: A union space containing available ingredients for recipe ideation.
- *Ingredient Set* $\mathbb{S}$: An finite subset $\mathbb{S} \in \mathbb{U}$ containing ingredients. In addition, $\mathbb{S}_n$ is a $n$-sized ingredient set where $|\mathbb{S}_n| = n$ and $n \geq 1$.
- *Additional Ingredient* $i_a$: A single ingredient to be added to current ingredient set where $i_a \in \mathbb{U}$.
- *Food Affinity Score* $y$: A score that quantifies the suitability of adding $i_a$ to $\mathbb{S}_n$ resulting $\mathbb{S}_{n+1} = \mathbb{S}_n \cup \{i_a\}$ where $y \in \mathbb{R}$.

PROBLEM 1 (RECIPE IDEATION TASK). *We define the objective of* **Recipe Ideation Task** *as finite steps of sequential ingredient set expansion. Each step involves expanding the ingredient set $\mathbb{S}_n$ by adding another ingredient $i_a$ which results in $\mathbb{S}_{n+1} = \mathbb{S}_n \cup \{i_a\}$.*

To solve the above problem, we introduce the following two tasks.

TASK 1 (FOOD AFFINITY SCORE PREDICTION). *Given an ingredient set $\mathbb{S}_n$ and additional ingredient $i_a$, RecipeMind $f$ predicts the food affinity score between $\mathbb{S}_n$ and $\{i_a\}$ through modeling $y = f(\mathbb{S}, i_a)$.*

TASK 2 (ADDITIONAL INGREDIENT RECOMMENDATION). *Given an ingredient set $\mathbb{S}_n$, all possible ingredients $i \in \mathbb{U} - \mathbb{S}_n$ and RecipeMind $f$, the recommended ingredient $i_a$ to be added to $\mathbb{S}_n$ is based on the top-ranked affinity score predictions made by RecipeMind.*

$$i_a = \arg\max_i f(\mathbb{S}_n, i) \quad (1)$$

To train our RecipeMind model, we constructed a large-scale dataset where each data instance is defined as $(\mathbb{S}_n, i_a, y)$. The data instances were built from the ingredient subset co-occurrences in the Reciptor dataset containing 507,834 recipes and 2,391 ingredients deemed as Ingredient Space $\mathbb{U}$ [27]. The food affinity scores were calculated based on *Significant PMI based on Document Count* and were applied to our recipe ideation task [15, 16].

We adopted the Set Transformer framework when developing the model architecture of RecipeMind [26]. To help RecipeMind jointly learn cross-relational features between ingredients in $\mathbb{S}_n$ and $i_a$, we developed Cascaded Set Transformer using *Pooling by Multihead Cross-Attention (PMX)*.

We evaluated RecipeMind's food affinity score prediction through baseline and ablation experiments with expanding ingredient subsets including unseen sizes in training set. We further analyzed the recommendation results and attention heatmaps after deploying RecipeMind in example recipe ideation scenarios to explore its understanding in recipe contexts.

As shown in Figure 1, RecipeMind encompasses from food pairing to recipe completion as it chooses the most suitable ingredient choices given *any* number of ingredients in current set. To the best of our knowledge, this work is the first attempt to introduce a data-driven approach that assists ingredient choices at *any* stage in recipe ideation. The major contributions of our work can be summarized as follows,

- We formulated a downstream task called recipe ideation which features and food affinity score prediction and additional ingredient recommendation.
- We created a large-scale dataset that contains affinity scores for each pair of $n$-sized ingredient set and additional ingredient.
- We developed RecipeMind utilizing Cascaded Set Transformers using Pooling by Multihead Cross-Attention.
- We empirically demonstrated RecipeMind's robustness in expanding set sizes through experiments and analyzed its understanding in recipe contexts throughout recipe ideation scenarios [1].

## 2 RELATED WORKS

### 2.1 Representation Learning for Recipes

Previous works that have introduced various deep learning approaches for improved representation learning of cooking recipes. Cross-modal or multimodal approaches incorporating recipe texts with images have been introduced where some focused on improving representation learning on recipes [6, 8, 32, 33, 37, 48, 50, 53, 54]. Others utilized these features to improve recipe retrieval tasks [7, 9, 29, 58]. Few works have attempted to apply cross-modal feature learning to recipe numeracy tasks such as predicting calories [28] or food ingredient amounts [18].

Meanwhile, as recipes can be expressed as sets of ingredient, recent works have proposed set representation learning methods to effectively learn recipe-related contextual features [24, 27]. Our work also proposes to apply set representation learning to recipe ideation since it is crucial to understand various recipe context and determine the optimal ingredient to be added for the next step.

### 2.2 Previous Approaches related to Recipe Ideation

One of the downstream tasks related to our work is food pairing which can be deemed as the fundamental form of recipe ideation. Computational methods for food pairing have been introduced including Kitchenette. These methods utilize statistical co-occurrences or chemical similarities of ingredient pairs [1, 39, 40]. Another downstream task related to our work is recipe completion [14, 17, 24]. These methods optimized data-driven models to predict ingredients given partial or nearly completed recipes. Our work encompasses both tasks as recipe ideation aims to guide users' choices on ingredient additions given different ingredient sets.

---
[1]The source code and demo web page for RecipeMind are open for public access. (https://github.com/dmis-lab/RecipeMind , https://recipemind.korea.ac.kr)

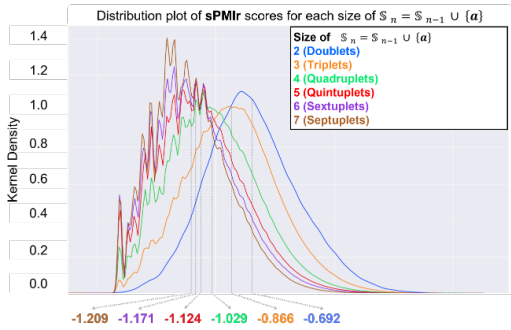

**Figure 2: Distribution of *sPMIr* scores for each size of $\mathbb{S}_n = \mathbb{S}_{n-1} \cup \{i_a\}$ where $n \geq 2$. The mean values of sPMIr-based affinity scores are shown in the x-axis for each size of ingredient subset. As *n* increases, the statistic mean of food affinity scores shifts toward negative while the overall distribution becomes non-normal.**

Other works have suggested approaches for generating recipes containing cooking instructions and/or ingredients. Various generative models using different modalities of queries such as food images, cooking videos and texts have been developed [19, 22, 43, 52, 57]. Recently, a system for recipe editing considering dietary constraints was introduced [10]. While recipe ideation and generation have common traits such as creativeness, RecipeMind guides users' ingredient choices step by step based on its predicted food affinity scores for all expanding sizes of ingredient sets.

## 3 DATASET
### 3.1 Obtaining n-sized Ingredient Subsets from Reciptor Dataset

Table 1 shows the dataset statistics involving *n*-sized ingredient subsets obtained from the Reciptor dataset. We extracted *n*-sized ingredient subsets from the Reciptor dataset containing 507,834 recipes which was originally used by Reciptor [27]. We adopted Kitchenette's approach by removing ingredients whose occurrence count does not exceed 20 [40]. As a result, we obtained 2,391 unique ingredients and used them to generate *n*-sized subsets based on their co-occurrence statistics in the Reciptor dataset [27].

While the number of possible 2-sized subsets (doublets) in our dataset is $\binom{2391}{2} = 2,857,245$, we adopted Kitchenette's approach by selecting doublets exceeding 5 occurrences in the dataset and obtained 236,297. The same criteria was applied to obtaining 3 and 4-sized subsets which resulted in a total of 1,226,767 among $\binom{2391}{3}$ and 1,952,345 among $\binom{2391}{4}$ possible subsets. Furthermore, we obtained all available 5,6 and 7-sized ingredient subsets for only testing purposes. Right side of Table 1 shows the obtained numbers of *n*-sized ingredient subsets.

### 3.2 Constructing Data Instances

For each *n*-sized ingredient subset obtained from the Reciptor dataset, we firstly split each *n*-sized ingredient set $\mathbb{S}_n$ into a pair of $\mathbb{S}_{n-1}$ and its additional ingredient $i_a$ where the total number of combinations is *n*. Therefore, the total number of data instances built from ingredient doublets, triplets and quadruplets is $236,297 \times 2$, $1,226,767 \times 3$ and $1,952,345 \times 4$ respectively. The doublet-based data instances are identical to ingredient pairings as the input ingredient set is a singleton. Left side of Table 1 shows the actual numbers of data instances contained in our dataset.

### 3.3 Generating sPMIr-based Food Affinity Scores

We generated the affinity scores for each data instances consisting a ingredient set and its additional ingredient. Kitchenette used *Normalized Point-wise Mutual Information* (NPMI) [5, 40, 49] to formulated food pairing scores. The scores represent the co-occurrence between two different ingredients and intuitively determine whether each ingredient pair is suitable or not.

In this work, we adopted Damani's *Significant PMI based on Document Count* (sPMId) score, an improved version of PMI considering statistical significance [15, 16]. The formulation of the sPMId score between words **x**, **y** is mathematically expressed as,

$$\text{sPMId}(\mathbf{x}, \mathbf{y}) = \log \frac{n(\mathbf{x}, \mathbf{y})}{\frac{n(\mathbf{x})n(\mathbf{y})}{N} + \sqrt{\max(n(\mathbf{x}), n(\mathbf{y}))} * \sqrt{\frac{\ln \delta}{-2.0}}} \quad (2)$$

where $n(\mathbf{x})$ is the number of documents that contain **x** at least once, *N* is the total number of documents and $\delta$ is the parameter varying between 0 and 1. Prior to applying the sPMId-based score formulation to our task, we substituted words with ingredient subsets of varying sizes.

We propose a modified approach compatible with ingredient subsets used in the Reciptor dataset [27]. The modifications are the following,

- The documents used to calculate occurrences are the recipes which contain a full list of ingredients used for cooking.
- Given two disjoint ingredient subsets **X** and **Y** ($|\mathbf{X} \cap \mathbf{Y}| = 0$), a co-occurrence measure is defined based on their union's occurrence ($\mathbf{X} \cup \mathbf{Y}$).
- As our task involves adding one ingredient to current to a *N*-sized ingredient set, we introduced *Significant PMI based on Recipe Count* (sPMIr) to formulate affinity scores for training RecipeMind using the above modifications.

The calculation of a sPMIr-based affinity score when adding an ingredient $i_a$ to *n*-sized ingredient set $\mathbb{S}_n$ is expressed as,

$$\text{sPMIr}(\mathbb{S}, i_a) = \log \frac{r(\mathbb{S}_{\bowtie} \cup \{i_a\})}{\frac{r(\mathbb{S}_n)r(\{i_a\})}{R} + \sqrt{\max(r(\mathbb{S}_n), r(\{i_a\}))} * \sqrt{\frac{\ln \delta}{-2.0}}} \quad (3)$$

where $r(\mathbf{X})$ is the number of recipes that used ingredient set **X** at least once, *R* is the total number of recipes, $\delta$ is set as 0.2 and $|\mathbb{S}_n \cap \{i_a\}| = 0$ since our task involves adding new ingredients. Unlike the pairing scores originally used in Kitchenette [40], the food affinity scores are not bounded since they are not normalized.

| Ingred. Subset Size | Number of | Number of | Number of Data Instances $(\mathbb{S}_{n-1}, i_a, y)$ | | | |
|---|---|---|---|---|---|---|
| n $(\mathbb{S}_n = \mathbb{S}_{n-1} \cup \{i_a\})$ | Possible Subsets $\binom{2391}{n}$ | Obtained Subsets $\mathbb{S}_n$ | Total | Training | Validation | Test |
| 2 $(\mathbb{S}_2 = \mathbb{S}_1 \cup \{i_a\})$ | 2.86e06 | 236,297 | 472,594 | 378,074 | 23,630 | 70,890 |
| 3 $(\mathbb{S}_3 = \mathbb{S}_2 \cup \{i_a\})$ | 2.28e09 | 1,226,767 | 3,680,301 | 2,944,239 | 184,017 | 552,045 |
| 4 $(\mathbb{S}_4 = \mathbb{S}_3 \cup \{i_a\})$ | 1.36e12 | 1,952,345 | 7,809,380 | 6,247,504 | 390,472 | 1,171,404 |
| 5 $(\mathbb{S}_5 = \mathbb{S}_4 \cup \{i_a\})$ | 6.48e14 | 1,567,562 | 7,837,810 | 0 | 0 | 7,837,810 |
| 6 $(\mathbb{S}_6 = \mathbb{S}_5 \cup \{i_a\})$ | 2.57e17 | 897,874 | 5,387,244 | 0 | 0 | 5,387,244 |
| 7 $(\mathbb{S}_7 = \mathbb{S}_6 \cup \{i_a\})$ | 8.78e19 | 439,348 | 3,075,436 | 0 | 0 | 3,075,436 |

Table 1: Statistics of RecipeMind dataset. The filtered 2391 ingredients were used to obtain $n$-sized ingredient subsets $\mathbb{S}_n$ used in the 507,834 recipes from the original Reciptor dataset. Only subsets whose ingredient co-occurrence count exceeds 5 were used. The obtained $n$-sized subsets were used to construct data instances for training RecipeMind where each consists a $(n-1)$-sized ingredient set $\mathbb{S}_{n-1}$, its additional ingredient $i_a$ and calculated food affinity score y. The data instances based on $\mathbb{S}_2, \mathbb{S}_3, \mathbb{S}_4$ were used for training, validation and testing while the remaining ones based on $\mathbb{S}_5, \mathbb{S}_6, \mathbb{S}_7$ were only used for testing.

| Ingredient Set Expansion | Ingredient Set $\mathbb{S}_n$ [#] | Additional Ingredient $i_a$ [#] | sPMIr for $\mathbb{S}_{n+1} = \mathbb{S}_n \cup \{i_a\}$ [#] |
|---|---|---|---|
| Adding *baking soda* to *flour, eggs* | flour, eggs [20205] | **baking soda [31840]** | **0.6887 [7510]** |
| | | vanilla [29857] | 0.6810 [6941] |
| | | nuts [5375] | 0.6521 [1393] |
| | | romaine lettuce [2009] | -1.5531 [6] |
| | | red wine vinegar [6999] | -1.5546 [12] |
| | | cucumbers [4196] | -1.5869 [6] |
| Adding *nuts* to *flour, eggs, baking soda* | flour, eggs, baking soda [7510] | **nuts [5375]** | **0.7582 [941]** |
| | | vanilla [29857] | 0.7565 [3296] |
| | | buttermilk [8217] | 0.7272 [1139] |
| | | parmesan cheese [29226] | -1.8916 [8] |
| | | garlic powder [21429] | -1.8993 [6] |
| | | garlic cloves [65879] | -2.1562 [9] |
| Adding $i_a$ to $\mathbb{S}_3$ given a fixed set $\mathbb{S}_4$ *flour, eggs, baking soda, nuts* | flour eggs nuts [1393] | baking soda [31840] | 0.5669 [941] |
| | flour nuts baking soda [1443] | eggs [77046] | 0.3849 [941] |
| | eggs nuts baking soda [1249] | flour [56501] | 0.3155 [941] |

Table 2: Examples of data instances from RecipeMind dataset with calculated scores and occurrence counts ([#]) in original dataset. The first two merged rows show ingredient set expansion examples based on calculated food affinity scores (sPMIr) where each example has a list of 3 highest- and lowest-scoring ingredients. For example, given *flour* and *eggs*, adding *baking soda* gives the highest score (0.6887) while adding *cucumbers* gives the lowest (-1.5869). The last merged row shows sPMIr scores being calculated differently based on the selected additional ingredient within ingredient set (*flour, eggs, baking soda, nuts*).

### 3.4 Preliminary Analysis on Generated Food Affinity Scores

Figure 2 shows the distribution of sPMIr scores for each size of $\mathbb{S}_{n+1} = \mathbb{S}_n \cup \{i_a\}$ where $n \geq 1$. The sPMIr-based food affinity scores calculated based on ingredient doublets ($n = 2$) and triplets ($n = 3$) show normal distributional behavior. However, the distribution tends to become skewed towards negative as the size increases. The shifting distribution of affinity scores may pose challenges to RecipeMind's generalization in expanding sets.

Table 2 shows the calculated sPMIr-based affinity scores for example ingredient combinations actually used in the Reciptor dataset. Given an ingredient set **flour** and **eggs**, the top scoring additional ingredients are **baking soda**, **vanilla** and **nuts**. These three ingredients are known to be popularly used with **flour** and **eggs** in baking recipes [41, 47].

Higher co-occurrence counts between $\mathbb{S}_n$ and $i_a$ have the tendency to result in higher affinity scores. For example, while **nuts**, **red wine vinegar** and **cucumbers** have similar occurrences in the dataset, adding **nuts** yields the highest affinity score among them due to its co-occurrence with **flour** and **eggs**. The same applies to adding **vanilla** compared to **parmesan cheese**, **garlic cloves** given **flour**, **eggs** and **baking soda**.

Moreover, adding relatively unpopular ingredients may be compensated with higher scores as long as their co-occurrence with current set of ingredients is relatively higher. For instance, while adding **vanilla** yields a higher affinity score than **nuts**, an expanded set added with **baking soda** yields an opposite order of affinity scores. While **nuts** have been used less than **vanilla** fourfold, their co-occurrence with **flour**, **eggs** and **baking soda** is rewarded by sPMIr-based score formulation.

The pairing scores for $\mathbb{S}_2$ are symmetric (i.e. adding **x** to {**y**} and vice versa results the same score). However, different combinations of $(\mathbb{S}_n, i_a), n > 2$ resulted in different food affinity scores as shown in the last 3 rows of table 2. These characteristics presented the necessity of understanding the inter-relational features between recipe context of $\mathbb{S}_n$ and $i_a$.

Inferring the missing relations between $\mathbb{S}_n$ and $i_a$ from $\mathbb{S}_{n+1} = \mathbb{S}_n \cup \{i_a\}$ is important in recipe ideation. We performed *ingredient subset*-based data split for partitioning the $n$-sized ingredients subsets ($n = 2, 3, 4$) into training, validation and test purposes (8:0.5:1.5). The validation data instances were used for searching the optimal hyper-parameters for RecipeMind. We ensured that data instances $(\mathbb{S}_{n-1}, i_a, \mathbf{y})$ in different partitions do not share the same ingredient subset $\mathbb{S}_n$. The remaining ingredient subsets ($n = 5, 6, 7$) were used for only testing purposes.

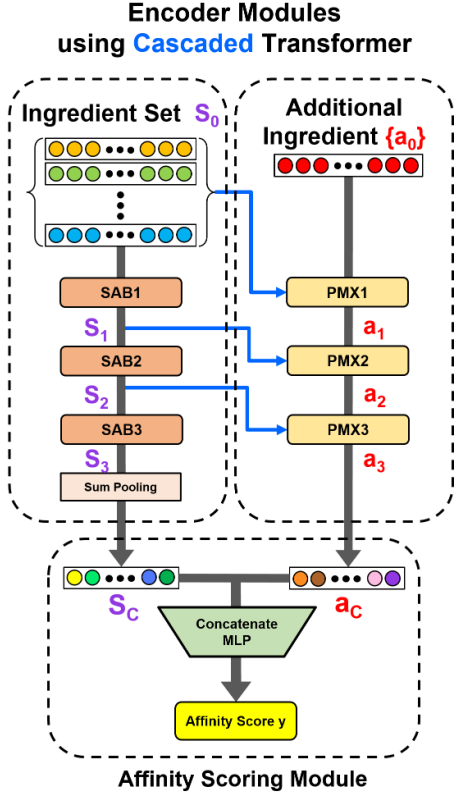

**Figure 3: Model Archtitecture of RecipeMind.** $S_0$, $a_0$ is the set of $n$ ingredients n and its additional ingredient $i_a$ both encoded ingredient-wise by a shared MLP. $S_l$ and $a_l$ are intermediate encoded representations while $S_C$ and $a_C$ are final contextual representations used to predict a food affinity score.

## 4 MODEL ARCHITECTURE OF RECIPEMIND

### 4.1 Overview

RecipeMind ($f(\mathbb{S}_n, a_i) = y$) takes a $n$-sized ingredient set $\mathbb{S}_n$ and an additional ingredient $i_a$ as input to predict the food affinity score of the updated $(n+1)$-sized set $\mathbb{S}_n \cup \{i_a\}$. It consists of three modules which are the Ingredient Set encoder, Additional Ingredient encoder and Affinity Scoring module. Figure 3 shows an overall description for RecipeMind's model architecture.

Prior to introducing the model architecture of RecipeMind, we refer to Kitchenette built on Siamese Neural Networks which predicts food pairing scores [40]. One of the main justifications for this model design choice is the homogeneity of two inputs (*ingredients*). Meanwhile, Reciptor and RecipeBowl employed Set Transformers effective representation learning of recipes [24, 27] where both models handle *ingredient sets*.

As the additional ingredient $i_a$ can be deemed as a 1-sized *ingredient set*, we may merge two model design choices. However, we propose an alternative approach by using *Cascaded Set Transformers with Pooling by Multihead Cross-Attention*.

### 4.2 Ingredient Set Encoder Module

The initial word-based representations of ingredients in $n$-sized set $\mathbb{S}_n$ are encoded to $S_0$ by a 2-layered element-wise multi-layer perceptron (MLP) shared in both encoder modules. We used 300-dimensional FlavorGraph embeddings previously trained based on chemical relationships between food ingredients and flavor compounds [39]. The dimension sizes for the encoded ingredients are uniformly set to $h = 128$.

The shared MLP that takes both representations $\mathbb{S}_{n+1} = \mathbb{S}_n \cup \{a_i\}$ ingredient-wise as input is mathematically expressed as,

$$\mathbf{H} = \sigma(\text{Dropout}(\text{Linear}_{shared1}(\mathbb{S}_{n+1}))) \quad (4)$$

$$\mathbb{S}_0 \cup \{\mathbf{a}_0\} = \sigma(\text{Dropout}(\text{Linear}_{shared2}(\mathbf{H}))) \quad (5)$$

where $S_0 \in \mathbb{R}^{n \times 128}$ and $a_0 \in \mathbb{R}^{128}$ are the $n$ 128-dimensional encoded ingredient embeddings in the current set and 128-dimensional additional ingredient embedding. The weights and bias in Linear$_{shared1}$ and Linear$_{shared2}$ are $W_{shared1} \in \mathbb{R}^{300 \times 128}$, $b_{shared1} \in \mathbb{R}^{128}$ and $W_{shared2} \in \mathbb{R}^{128 \times 128}$, $b_{shared2} \in \mathbb{R}^{128}$ respectively. $\sigma$ is the Rectified Linear Unit (ReLU) activation function while Dropout is dropout layer with probability of 0.025.

The encoded ingredients in current set are propagated through 3 stacked Set Attention Blocks (SAB) followed by Sum Pooling [26]. Each successive SAB uses self-attention mechanism to encode higher order ingredient-ingredient relations and form recipe context based on the set of ingredients [24].

The Ingredient Set encoder in RecipeMind taking an $n$-sized encoded ingredient set $\mathbb{S}_n$ as input is mathematically expressed as

$$S_0 = \text{MLP}_{shared}(\mathbb{S}_n) \quad (6)$$

$$S_l = \text{SAB}_l(S_{l-1})(l = 1, 2, 3) \quad (7)$$

$$S_C = \text{SumPool}(S_3) \quad (8)$$

$S_l \in \mathbb{R}^{n \times 128}$ is a set of $n$ 128-dimensional ingredient embbedings encoded by the $l$th SAB ($\text{SAB}_l$). $\mathbb{S}_C \in \mathbb{R}^{128}$ is a 128-dimensional contextualized embedding for the current ingredient set. SumPool is a permutation-invariant sum pooling operator.

Each SAB is defined as a Multihead Attention Block (MAB) using the same elements as query, key and value [26, 51] and applies self-attention to a set of elements. The SABs followed by the Sum Pooling operator of the Recipe Idea Encoding Layer are mathematically expressed as,

$$\text{SAB}_l(S_{l-1}) = \text{MAB}_l(S_{l-1}, S_{l-1}) \quad (9)$$

$$\text{MAB}(\mathbf{X}, \mathbf{Y}) = \text{LayerNorm}(\mathbf{H} + \text{RFF}_1(\mathbf{X})) \quad (10)$$

$$\mathbf{H} = \text{LayerNorm}(\mathbf{X} + \text{RFF}_2(\text{MultiAttn}(\mathbf{X}, \mathbf{Y}, \mathbf{Y}))) \quad (11)$$

$\text{SAB}_l$ is the $l$th Set Attention Block, $\text{MAB}_l$ is the $l$th Multihead Attention Block, LayerNorm is layer-wise normalization [3] and RFF is row-wise feedforward layer consisting three consecutive MLPs without Dropout layer and using ReLU as non-linear activation function. MultiAttn is an attention mechanism module with 8 heads [51] where the attention weights in each head given query vectors $\mathbf{Q} \in \mathbb{R}^{n_q \times 128}$, key vectors $\mathbf{K} \in \mathbb{R}^{n_k \times 128}$ and value vectors $\mathbf{V} \in \mathbb{R}^{n_k \times 128}$ are calculated.

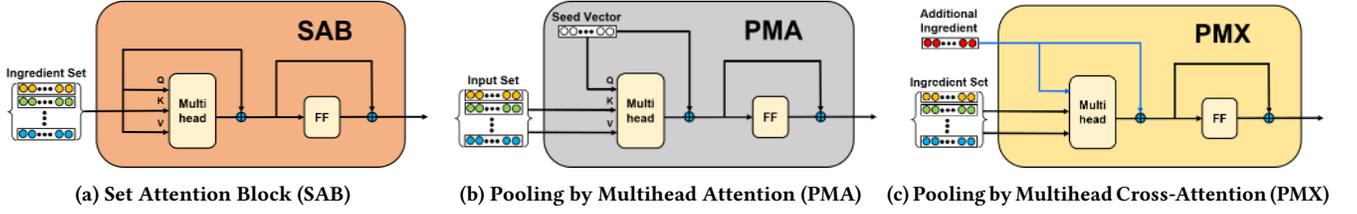

(a) Set Attention Block (SAB)    (b) Pooling by Multihead Attention (PMA)    (c) Pooling by Multihead Cross-Attention (PMX)

Figure 4: Attention blocks used in Set Transformer framework and RecipeMind. The SAB (a) encodes the ingredients using self-attention. The PMA (b) aggregates the set of elements using trainable seed vector. The PMX (c) used in RecipeMind's Cascaded Transformer encodes the additional ingredient based on its cross-attention against set of ingredients.

The aggregation of ingredient-wise representations $S_3$ from the last SAB is mathematically expressed as follows,

$$S_C = \text{SumPool}(S_3) \quad (12)$$

$$\text{SumPool}(S_3) = \sum_{j=0}^{|S_3|} i_j \quad (13)$$

where SumPool is a permutation-invariant Sum Pooling operator that performs element-wise summation of the ingredient representations $i_j \in S_3$.

### 4.3 Additional Ingredient Encoder Module

The authors of Set Transformer framework introduced Pooling by Multihead Attention (PMA) as illustrated in Figure 4b [26]. In this block, element-wise representations in set are aggregated by attending them to a single trainable seed vector. Adopting the methods proposed in PercieverIO and PICASO [23, 56], we devised Pooling by Multihead Cross-Attention (PMX) to improve representation learning on $i_a$ in different recipe contexts. As shown in Figure 4c, PMX is a variant of PMA where a set of items is aggregated based on multihead attention applied on another item instead of the seed vector [26].

The additional ingredient is refined through the shared MLP and 3 successive PMX blocks where each intermediate representation $a_l$ from the $l$th *PMX* block and its corresponding set from the $l$th SAB in Ingredient Set encoder is fed to the next $l + 1$th PMX layer. We denote this as *Cascaded Set Transformer* since Ingredient Set and Additional Ingredient encoder are jointly connected.

The Additional Ingredient Encoder module taking the additional ingredient $i_a$ as input is mathematically expressed as,

$$a_0 = \text{MLP}_{shared}(\{i_a\}) \quad (14)$$

$$a_l = \text{PMX}_l(a_{l-1}, S_{l-1})(l = 1, 2, 3) \quad (15)$$

$$a_C = a_3 \quad (16)$$

$a_l \in \mathbb{R}^{128}$ is a 128-dimensional ingredient embedding encoded by the $l$th PMX block ($\text{PMX}_l$) based on cross-attention between $a_{l-1}$ and $S_{l-1}$. $a_C \in \mathbb{R}^{128}$ is a 128-dimensional contextualized embedding for the additional ingredient.

The $l$th $PMX_l$ block that calculates cross-attention between $S_{l-1}$ and $a_{l-1}$, and outputs the refined representation $a_l$ is mathematically expressed as,

$$\text{PMX}_l(a_{l-1}, S_{l-1}) = \text{MAB}_l(a_{l-1}, S_{l-1}) \quad (17)$$

where the $MAB_l$ is $l$th Multihead Attention Block that computes the attention weights using $a_{l-1}$ as query vector and $S_{l-1}$ as key and value vectors.

### 4.4 Affinity Scoring Module

The Affinity Scoring module concatenates the final contextual representations vector-wise from both sides of encoding layers ($S_C, a_C$) to predict the ideation score $\hat{y}$ of adding an ingredient to current recipe idea and is mathematically expressed as follows,

$$\hat{y} = \text{MLP}_{score}(S_C \oplus a_C) \quad (18)$$

$\oplus$ is vector-wise concatenation between two contextualized embeddings. $\text{MLP}_{score}$ is a 2-layered MLP for predicting the food affinity score where only the intermediate layer uses Dropout and ReLU as non-linear activation. The weights, bias in the first and second linear layer of $\text{MLP}_{score}$ are $W_{score1} \in \mathbb{R}^{256 \times 128}, b_{score1} \in \mathbb{R}^{128}$ and $W_{score2} \in \mathbb{R}^{128 \times 1}, b_{score2} \in \mathbb{R}^1$ respectively.

### 4.5 Model Training

We used root mean squared error as loss objective for training RecipeMind on predicting the food affinity score $y$ for given $\mathbb{S}_n$ and $a_i$. The optimizer used in the training process is Adam [25] where the learning rate and weight decay is set to 1e-4, 1e-5 respectively. All RecipeMind and its ablations were trained to a maximum of 30 epochs with batch size of 1024 and early stopping.

## 5 EXPERIMENTS AND ANALYSIS

### 5.1 Model Baseline and Ablations

We conducted experiments to evaluate and compare our proposed RecipeMind's performance on affinity score prediction with other baselines and ablations. We firstly made naive predictions based on the statistic mean or median of all affinity scores in training set.

Since food affinity score prediction is a newly formulated task, we borrowed the architecture used in Kitchenette, Reciptor and RecipeBowl [24, 27, 40]. As Kitchenette can only predict affinity scores when adding an ingredient to a 1-sized ingredient set ($|\mathbb{S}_1 \cup \{i_a\}| = 2$), we only trained and tested it on data instances only based on doublets. In context of RecipeMind, both the Ingredient Set and Additional Encoder Module have weight-sharing MLPs while the Ideation Scoring Module has Wide-and-Deep layer [11].

For comparison involving larger ingredient sets, we replaced the weight-sharing MLPs with the weight-sharing Set Transformers used in Reciptor and RecipeBowl. Unlike Kitchenette, the Affinity

### (a) Evaluation results of main experiments using RMSE↓.

| Expanding Ingredient Set | | 2 | 3 | 4 | 5 | 6 | 7 |
|---|---|---|---|---|---|---|---|
| Naïve Guessing | mean value (-0.9661) | 0.473 | 0.4029 | 0.3868 | 0.3998 | 0.4076 | 0.4088 |
| | median value (-0.9830) | 0.4829 | 0.4073 | 0.3843 | 0.3935 | 0.3994 | 0.3991 |
| | **RecipeMind** | 0.2285 (0.0030) | 0.1560 (0.0019) | 0.1338 (0.0012) | 0.1307 (0.0036) | 0.1493 (0.0116) | 0.1821 (0.0249) |
| Baseline Models | Reciptor (Li et al.) | 0.2285 (0.0031) | 0.1662 (0.0011) | 0.1443 (0.0008) | 0.4830 (0.0743) | 0.6810 (0.0443) | 0.8837 (0.0781) |
| | RecipeBowl (Kim et al.) | 0.2413 (0.0033) | 0.1782 (0.0021) | 0.1566 (0.0023) | 0.4626 (0.1060) | 0.6469 (0.0530) | 0.8646 (0.0776) |
| | Kitchenette (Park et al.) | 0.2247 (0.0017) | | | | | |

### (b) Evaluation results of main experiments using PCORR↑.

| Expanding Ingredient Set | | 2 | 3 | 4 | 5 | 6 | 7 |
|---|---|---|---|---|---|---|---|
| Naïve Guessing | mean value (-0.9661) | | | | | | |
| | median value (-0.9830) | | | | | | |
| | **RecipeMind** | 0.8048 (0.0060) | 0.9170 (0.0023) | 0.9374 (0.0014) | 0.9384 (0.0023) | 0.9207 (0.0069) | 0.8819 (0.0160) |
| Baseline Models | Reciptor (Li et al.) | 0.8047 (0.0059) | 0.9055 (0.0013) | 0.9260 (0.0010) | 0.7044 (0.0789) | 0.6543 (0.0273) | 0.4848 (0.0332) |
| | RecipeBowl (Kim et al.) | 0.7790 (0.0066) | 0.8901 (0.0027) | 0.9120 (0.0028) | 0.6900 (0.0922) | 0.6314 (0.0291) | 0.4640 (0.0528) |
| | Kitchenette (Park et al.) | 0.6631 (0.0032) | | | | | |

Figure 5: Evaluation results of baseline experiments. The columns represent the expanded set size $n$ after adding an ingredient ($\mathbb{S}_n = \mathbb{S}_{n-1} \cup \{i_a\}$) while colored ones indicate unseen ingredient set sizes. Darker colors in cells indicate better results.

### (a) Evaluation results of ablation experiments using RMSE↓.

| Expanding Ingredient Set | | 2 | 3 | 4 | 5 | 6 | 7 |
|---|---|---|---|---|---|---|---|
| Ablations on Ingredient Set Encoder | **RecipeMind** | 0.2285 (0.0030) | 0.1560 (0.0019) | 0.1338 (0.0012) | 0.1307 (0.0036) | 0.1493 (0.0116) | 0.1821 (0.0249) |
| | RecipeMind w/o Cascaded PMX | 0.2346 (0.0034) | 0.1698 (0.0027) | 0.1473 (0.0020) | 0.1423 (0.0022) | 0.1578 (0.0056) | 0.1859 (0.0112) |
| | Rep the Set (Skianis et al.) | 0.2502 (0.0007) | 0.1837 (0.0010) | 0.1653 (0.0020) | 0.1729 (0.0083) | 0.2437 (0.0426) | 0.3558 (0.0910) |
| | Deep Sets (Zaheer et al.) | 0.2670 (0.0035) | 0.1945 (0.0012) | 0.1715 (0.0009) | 0.1653 (0.0017) | 0.1825 (0.0087) | 0.2153 (0.0213) |
| Ablations on Set Pooling Method | PMA (Lee et al.) | 0.2434 (0.0293) | 0.1582 (0.0031) | 0.1344 (0.0021) | 0.3345 (0.0421) | 0.5625 (0.0653) | 0.7439 (0.0865) |
| | Mean Pooling | 0.3098 (0.0479) | 0.2758 (0.0023) | 0.1333 (0.0011) | 0.4173 (0.1330) | 0.6221 (0.0219) | 0.8112 (0.0272) |
| | Max Pooling | 0.2298 (0.0023) | 0.1588 (0.0010) | 0.1353 (0.0004) | 0.2399 (0.0696) | 0.3829 (0.0655) | 0.5082 (0.0579) |

### (b) Evaluation results of ablation experiments using PCORR↑.

| Expanding Ingredient Set | | 2 | 3 | 4 | 5 | 6 | 7 |
|---|---|---|---|---|---|---|---|
| Ablations on Ingredient Set Encoder | **RecipeMind** | 0.8048 (0.0060) | 0.9170 (0.0023) | 0.9374 (0.0014) | 0.9384 (0.0023) | 0.9207 (0.0069) | 0.8819 (0.0160) |
| | RecipeMind w/o Cascaded PMX | 0.7926 (0.0066) | 0.9007 (0.0034) | 0.9226 (0.0024) | 0.9232 (0.0023) | 0.9002 (0.0045) | 0.8493 (0.0136) |
| | Rep the Set (Skianis et al.) | 0.7608 (0.0014) | 0.8850 (0.0019) | 0.9055 (0.0018) | 0.8945 (0.0096) | 0.8370 (0.0340) | 0.7645 (0.0486) |
| | Deep Sets (Zaheer et al.) | 0.7222 (0.0074) | 0.8677 (0.0017) | 0.8935 (0.0013) | 0.8939 (0.0018) | 0.8648 (0.0096) | 0.8097 (0.0186) |
| Ablations on Set Pooling Method | PMA (Lee et al.) | 0.7722 (0.0653) | 0.9145 (0.0035) | 0.9360 (0.0022) | 0.8516 (0.0272) | 0.7398 (0.0273) | 0.6197 (0.0351) |
| | Mean Pooling | 0.6493 (0.0842) | 0.8207 (0.0031) | 0.9375 (0.0011) | 0.8315 (0.0643) | 0.7323 (0.0143) | 0.6103 (0.0150) |
| | Max Pooling | 0.8027 (0.0042) | 0.9137 (0.0011) | 0.9353 (0.0004) | 0.8644 (0.0610) | 0.7715 (0.0553) | 0.6719 (0.0508) |

Figure 6: Evaluation results of ablation experiments.

Scoring Module has the same structure as RecipeMind's. These baseline models along with RecipeMind were trained and tested on data instances based on 2,3,4-sized subsets ($|\mathbb{S} \cup \{i_a\}| = 2, 3, 4$). Furthermore, we performed zero-shot testing on unseen ingredient set sizes ($|\mathbb{S} \cup \{i_a\}| = 5, 6, 7$).

The models for our baseline experiment are summarized as the following,

- **Naive guessing by statistics (mean, median)**
- **Kitchenette**
- **Reciptor:** Previously used for recipe representation learning, this set transformer consists 2 Induced Set Attention Blocks (ISABs) using 4 attention heads in their MABs followed by Pooling by Multihead Attention using 2 seed vectors [27]. We denote this as Reciptor for brevity.
- **RecipeBowl:** Previously used for recipe completion task, this set transformer consists 1 ISAB using 2 attention heads followed by Pooling by Multihead Attention using 1 seed vector [24]. We denote this as RecipeBowl for brevity.

We also conducted ablation experiments on RecipeMind's Encoder module and Set Pooling method in Ingredient Set Encoder module. The model ablations for RecipeMind's Encoder Modules are the following,

- **- PMX, + weight-sharing SABs:** We performed this experiment to assess the benefits of using larger weight-sharing Set Transformers. Note that RecipeMind's Set Transformer is larger than Reciptor's or Recipebowl's. We denote this as RecipeMind w/o Cascaded PMX for brevity.
- **- PMX, - SAB, + Rep the Set:** We replaced the Set Transformer related blocks with Rep the Set [45]. Since the original implementation of Rep the Set incurs heavily computational costs, we used an approximated version of it instead. We denote this as Rep the Set for brevity.
- **- PMX, - SAB:** We performed this experiment to verify employing deeper permutation-invariant models. This ablated model aligns with Deep Sets [55] as it only uses MLP for encoding elements. We denote this as Deep Sets for brevity.

| Step | Ingredient Set | Top 3 Recommendations by RecipeMind | | |
|---|---|---|---|---|
| 1 | **carrots, onions** | **celery** | bay leaves | potatoes |
| 2 | + celery | **potatoes** | bay leaves | cabbage |
| 3 | + potatoes | **cabbage** | beef stew meat | bouillon |
| 4 | + cabbage | **cabbage heads** | tzatziki | canning salt |
| 5 | + cabbage heads | **bouillon** | tzatziki | stout beer |
| 6 | + bouillon | **seitan** | stout beer | chicken tenderloins |
| 7 | + seitan | **chicken tenderloins** | string beans | tzatziki |
| 8 | + chicken tenderloins | **tzatziki** | string beans | ditalini |

Figure 7: Ingredient recommendations yielded by RecipeMind starting with carrots and onions. Top 3 recommendations are based on their predicted food affinity scores when being added to current ingredient set. Red colors indicate the initial set of ingredients given to RecipeMind and first step. The final set of ingredients implicate RecipeMind has formed recipe context related to soups and stews.

| Step | Ingredient Set | Top 3 Recommendations by RecipeMind | | |
|---|---|---|---|---|
| 1 | **buttermilk, flour** | **baking soda** | vanilla | baking powder |
| 2 | + baking soda | **baking powder** | vanilla | cocoa |
| 3 | + baking powder | **eggs** | vanilla | shortening |
| 4 | + eggs | **vanilla** | cocoa | shortening |
| 5 | + vanilla | **cocoa** | shortening | sour milk |
| 6 | + cocoa | **sour milk** | Crisco | vanilla flavoring |
| 7 | + sour milk | **baking chocolate** | apple jelly | cherry juice |
| 8 | + baking chocolate | **cherry juice** | candied fruit | apple jelly |

Figure 8: Ingredient recommendations yielded by RecipeMind starting with buttermilk and flour. The final set of ingredients implicate RecipeMind has formed recipe context related to bakery products such as cakes.

## 5.2 Experimental Results on Food Affinity Score Prediction

All experiments were conducted on the same training-validation-test split on data instances. We trained and tested the model with different random seeds 5 times and calculated the mean and standard deviation of evaluation metrics which are root mean squared error (RMSE) and Pearson's correlation (PCORR).

Figure 5 shows the evaluation results of baseline experiments. As shown in Figure 5 RecipeMind outperformed in almost all expanding sizes of ingredient sets in both RMSE and PCORR while Kitchenette achieved the best RMSE among other models given $|\mathbb{S}_{n-1} \cup \{i_a\}| = 2$. While Receptor and RecipeBowl achieved results for $|\mathbb{S}_{n-1} \cup \{i_a\}| = 2, 3, 4$ sub par with RecipeMind, they failed to demonstrate generalization in unseen set sizes $|\mathbb{S}_n \cup \{i_a\}| = 5, 6, 7$, falling far behind the naive guessing baselines.

The ablation results shown in Figure 6 further support our design choices for RecipeMind's Encoder Modules and Set Pooling in its Ingredient Set Encoder module. Comparing with ablations on Set Encoder modules, RecipeMind using Cascaded Set Transformers with PMX showed better results than other Encoder ablations. Comparing with ablations on Set Pooling methods, while PMA or Max Pooling showed similar performance when $|\mathbb{S}_{n-1} \cup \{i_a\}| = 2, 3, 4$, both ablations failed to generalize when $|\mathbb{S}_{n-1} \cup \{i_a\}| = 5, 6, 7$ which supports our choice of Sum Pooling. Overall, the baseline and ablation evaluative results show RecipeMind's predictability of food affinity scores in expanding ingredient set sizes including unseen ones ($n = 5, 6, 7$) in the training dataset.

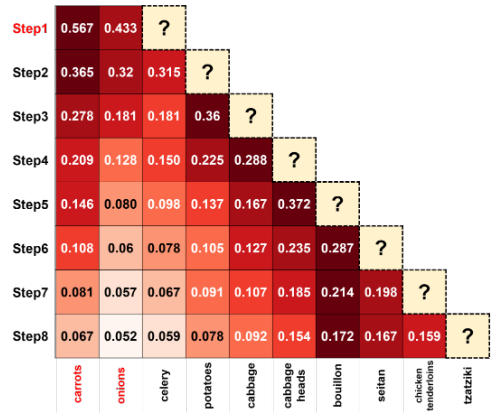

Figure 9: Attention weights extracted from RecipeMind's ideation scenario (Figure 7).

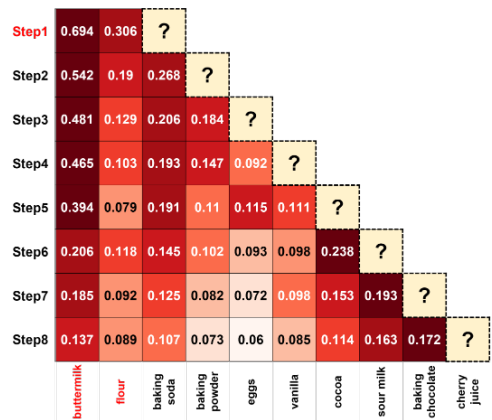

Figure 10: Attention weights extracted from RecipeMind's ideation scenario (Figure 8).

## 5.3 Qualitative Analysis on RecipeMind

### 5.3.1 Methods

Throughout recipe ideation, recipe context may change continuously due to different ingredient choices and combinatory features in set of ingredients. To investigate this, we deployed RecipeMind in two recipe ideation scenarios where (**carrots**, **onions**) and (**buttermilk**, **flour**) are given as initial set of ingredients $\mathbb{S}_2$. In each set expansion step of ideation scenario, RecipeMind predicts food affinity scores for all available ingredients $i \in \mathbb{U} - \mathbb{S}_n$ being added to $\mathbb{S}_n$. The highest scoring ingredient is selected and added to current set ($i_a = \arg\max_i f(\mathbb{S}_n, i)$, $\mathbb{S}_{n+1} = \mathbb{S}_n \cup \{i_a\}$). We repeated this step 8 times yielding a total of 10 ingredients.

Figures 7 and 8 show the top 3 ingredient recommendations for each step of set expansion in RecipeMind's ideation scenario. For example in Figure 7 at Step 1, RecipeMind given **carrots** and **onions** predicted the highest food affinity score for **celery**. Subsequently, **celery** is added to current set which is given to RecipeMind for retrieving another ranked list of ingredient recommendations (**potatoes**, **bay leaves**, **cabbage**) at Step 2.

Figures 9 and 10 show the cross-attention weights represented in heatmaps. The attention weights between $n$-sized current ingredient set $\mathbb{S}_n$ and additional ingredient $i_a$ were extracted from RecipeMind's last PMX block and averaged head-wise [12, 42]. The $p$th row and $q$th column represent the $p$th ingredient set expansion step and $q$th ingredient in current set respectively. For example at Step 1, when top-scoring **celery** was added to current set of **carrots** and **onions**, the cross-attention weights of these two ingredients were calculated as 0.567 and 0.432 respectively. At Step 2, when top-scoring **potatoes** was added, the cross-attention weights were calculated as 0.365, 0.32 for **carrots** and **onions** respectively and 0.315 for the previously added **celery**. The cell colors represent the ingredient's rank of attention weight (i.e the higher, the darker).

#### 5.3.2 Case Study 1: Narrowing Recipe Context

In the first ideation scenario shown in Figures 7 and 9, RecipeMind was given an initial set of two vegetables **carrots** and **onions** which are used in a variety of dishes such *soups & stews*, *salads* or *beef* recipes [31, 34, 46]. The attention weights in less skewed distribution from Step 1 (0.567, 0.433) to Step 2 (0.365, 0.320, 0.315) show that RecipeMind hasn't fully determined one of the 3 recipe contexts.

From Step 3 to Step 5, while the top 3 recommendations are diverse in ingredient categories, RecipeMind predicted higher affinity scores for adding vegetable ingredients. Previously added ingredients **potatoes** (0.360), **cabbage** (0.288), **cabbage heads** (0.372) were consecutively assigned with highest attention weights. We speculated that RecipeMind focused on common ingredient characteristics in determining food affinities and narrowed its recipe context to either *soups & stews* or *salads*.

After being added to the ingredient set at Step 6, **bouillon** which is the core ingredient of soups [20], has remained dominantly attentive throughout the rest of ideation (0.287, 0.214, 0.172). Follow-up additions **seitan** and **chicken tenderloins** are possible ingredient alternatives for cooking *soups & stews* dishes [2?]. As RecipeMind has further narrowed down its recipe context to *soups & stews*, we expect future ingredient additions to be highly relevant to it.

#### 5.3.3 Case Study 2: Main and Supportive Recipe Context

In the first ideation scenario shown in Figures 8 and 10, RecipeMind was given initial set of **buttermilk** and **flour** as they are essential ingredients for *bakery* recipes [4, 35]. From Step 1 to Step 5, RecipeMind suggested more essential ingredients (**baking soda**, **baking powder**, **eggs**, **vanilla**) while maintaining its main *bakery* recipe context based on **buttermilk**.

Throughout ideation, the attention weights assigned to **buttermilk** did not relatively decay as much as the initial ingredients in first scenario (0.694, 0.542, ... , 0.185, 0.137). The top 3 recommendations from Step 6 to Step 8 were mostly supportive ingredients such as **cherry juice** (beverage) and **apple jelly** (sweet sauces). Moreover, the previously added ingredients **cocoa** (0.238), **sour milk** (0.193), **baking chocolate** (0.172) were consecutively assigned with highest attention weights, implicating their own supportive recipe context. Merging these two recipe contexts will likely lead to chocolate cake recipes [?] served with **cherry juice** recommended at Step 8.

#### 5.3.4 Summary

We deployed RecipeMind in two recipe ideation scenarios to examine its ingredient set expansion process and interpret the calculated cross-attention weights. Our analysis demonstrate different ingredient choices in recipe ideation lead to different recipe contexts and completion. The attention weights from each the two ideation scenarios show different changes in RecipeMind's understanding of recipe context. However, we speculate that RecipeMind is open to vast possibilities of recipe ideation and its completion depending on users' choices on ingredients.

## 6 CONCLUSION AND FUTURE WORK

We devised a computational approach for recipe ideation by proposing two tasks; food affinity score prediction and additional ingredient recommendation. We implemented RecipeMind using the Cascaded Set Transformer to help it jointly learn features between current ingredient set and its additional ingredient. We then trained it on our constructed dataset containing food affinity scores. Experimental results including ablations demonstrate RecipeMind's robustness in predicting affinity scores for expanding ingredient sets. Qualitative analysis provides insight in how RecipeMind understands recipe context in set of ingredients.

While our definition of recipe ideation is confined to adding ingredients to current set, we may expand this into combining two $n$, $m$-sized ingredient sets, creating different combinations of recipe context and deriving novel recipe ideas in the end. Since the food affinity scores are mainly based on co-occurrence statistics, our next step is improving RecipeMind with other important food-related aspects such as nutrition and flavor chemistry. We plan to add nutrition constraints for addressing health benefits and incorporate prior knowledge on flavor chemistry to enhance RecipeMind's robustness.

Our work is part of a collaboration with Sony AI and their Gastronomy Flagship Project, and the aim is to deploy RecipeMind in food-related applications in order to interactively help chefs create delicious, healthy and sustainable dishes by uncovering vast new opportunities in ingredient combination. We expect RecipeMind to benefit the cuisine domain and accelerate food industry development in the future.

## ACKNOWLEDGEMENTS

This research was supported by the National Research Foundation of Korea (NRF-2020R1A2C3010638), the MSIT(Ministry of Science and ICT), Korea, under the ICT Creative Consilience program(IITP-2021-2020-0-01819) supervised by the IITP(Institute for Information & Communications Technology Planning & Evaluation) and Sony AI (https://ai.sony)